\documentclass[useAMS,usenatbib]{mn2e}
\usepackage{amsmath,graphicx,natbib}
\newcommand{\changed}[1]{{#1}}

\newcommand{\comment}[1]{}

\ifx\etal\undefined
	\newcommand{\etal}{{\it et al.}}
\fi

\title[Atmospheric Scintillation in Astronomical Photometry]{Atmospheric Scintillation in Astronomical Photometry}	
\author[J. Osborn et al.]{J. Osborn$^{1}$\thanks{E-mail:james.osborn@durham.ac.uk (JO)}, D. F\"ohring$^{1}$, V.~S. Dhillon$^{2}$ and R.~W. Wilson$^{1}$\\
$^{1}$Department of Physics, Centre for Advanced Instrumentation, University of Durham, South Road, Durham DH1 3LE, UK \\
$^{2}$Department  of Physics and Astronomy, University of Sheffield, Sheffield, S3 7RH, UK}

\date{\today}	
	
\begin{document}	

\maketitle		

\begin{abstract}
Scintillation noise due to the Earth's turbulent atmosphere can be a dominant noise source in high-precision astronomical photometry when observing bright targets from the ground. Here we describe the phenomenon of scintillation from its physical origins to its effect on photometry. We show that Young's (1967) scintillation-noise approximation used by many astronomers tends to underestimate the median scintillation noise at several major observatories around the world. We show that using median atmospheric optical turbulence profiles, which are now available for most sites, provides a better estimate of the expected scintillation noise and that real-time turbulence profiles can be used to precisely characterise the scintillation noise component of contemporaneous photometric measurements. This will enable a better understanding and calibration of photometric noise sources and the effectiveness of scintillation correction techniques. We also provide new equations for calculating scintillation noise, including for extremely large telescopes where the scintillation noise will actually be lower than previously thought. These equations highlight the fact that scintillation noise and shot noise have the same dependence on exposure time and so if an observation is scintillation limited, it will be scintillation limited for all exposure times. The ratio of scintillation noise to shot noise is also only weakly dependent on telescope diameter and so a bigger telescope may not yield a reduction in fractional scintillation noise.
\end{abstract}

\section{Introduction}

High-precision photometry is key to several branches of astronomical research, including (but not limited to) the study of extrasolar planets, astroseismology and the detection of small Kuiper-belt objects within our Solar System. The difficulty with such observations is that, although the targets are bright, the variations one needs to detect are often small (typically $\sim$0.01\% to $\sim$0.1\%). This is within the capabilities of modern detectors. However, when the light from the star passes through the Earth's atmosphere, regions of turbulence cause intensity fluctuations (seen as twinkling by the naked eye) called scintillation. This scintillation, which induces photometric variations in the range of $\sim$0.1\% to 1.0\%, limits the detection capabilities of ground based telescopes (e.g. \citealp{Brown94, Heasley96, Ryan1998}).

Knowing the level of scintillation noise is important because it will enable performance assessment, calibration and optimisation of photometric instrumentation. It will also help to explain and constrain model fits to photometric data (for example, extrasolar planet transit/eclipse light curves; \citealp{FohringThesis14}), and to help develop scintillation correction concepts such as Conjugate-Plane Photometry \citep{Osborn11}, Tomographic wavefront reconstruction \citep{Osborn15} and active deformable mirror techniques \citep{Viotto12}. It would also enable passive techniques such as `lucky photometry' where only data taken during photometric conditions (i.e. in times of low scintillation noise) are used in the reduction process.
 
Young proposed an equation which can be used to estimate the scintillation noise for an observation given the telescope's altitude and diameter, and the observation's exposure time and airmass. This equation is regularly used by many astronomers (for example \citealp{southworth2009}) to estimate the scintillation noise in their measurements. Recent work by \cite{Kornilov12d} showed that this equation tends to underestimate the median scintillation noise by a mean factor of 1.5.

As well as presenting new results on scintillation, we hope this paper will serve as a useful guide for astronomers to understand, estimate the size of, and correct for scintillation. We show that using the theoretical scintillation noise calculated from the median optical turbulence profile for a particular site is a better estimate of the median scintillation noise. However, we also show that the measured scintillation noise has large variability due to changes in the optical turbulence profile. Therefore, if one wishes to know the scintillation noise during a particular observation, the contemporaneous optical turbulence profile should be used if possible. With the proliferation of Adaptive Optics (AO) systems, real-time turbulence profilers are becoming more common-place at major observatories (see section~\ref{sect:prof}). As the scintillation noise is predominantly caused by high altitude turbulence we can assume that a single profiler at an observatory site can be used to precisely predict the scintillation noise applicable to all of the local telescopes. This dominance of the high altitude turbulence also explains why it is possible to observe highly photometric conditions during periods of bad seeing.

In section~\ref{sect:theory} the physical origin of scintillation noise is described and simulated light curves are shown. Section~\ref{sect:Young} reviews Young's scintillation approximation and in section~\ref{sect:prof} we compare the scintillation noise from Young's approximation with scintillation noise measurements from atmospheric profiling instruments. In section~\ref{sect:meas} we look at the magnitude of the scintillation noise on astronomical photometric measurements. In section~\ref{sect:largeTel} we present modified equations for the case of extremely large telescopes, when the outer scale of the turbulence becomes critical and, in section~\ref{sect:secObs}, telescopes with large secondary mirrors.

\section{Theory}
\label{sect:theory}
Optical turbulence is caused by the mechanical mixing of layers of air with different temperatures and hence density. The refractive index of air depends on its density and so  turbulence at a boundary of air masses with different temperatures creates a continuous screen of spatially and temporally varying refractive indices. 

The wavefront from an astronomical source can be considered flat at the top of the Earth's atmosphere. As it propagates to the ground it becomes aberrated by the optical turbulence which forms a limit to the precision of measurements from ground-based telescopes(for example, see figure 1 in \citealp{Osborn11}).

\begin{figure}
$\begin{array}{cc}
	\centering
	\hspace{-1mm}\includegraphics[width=0.5\columnwidth]{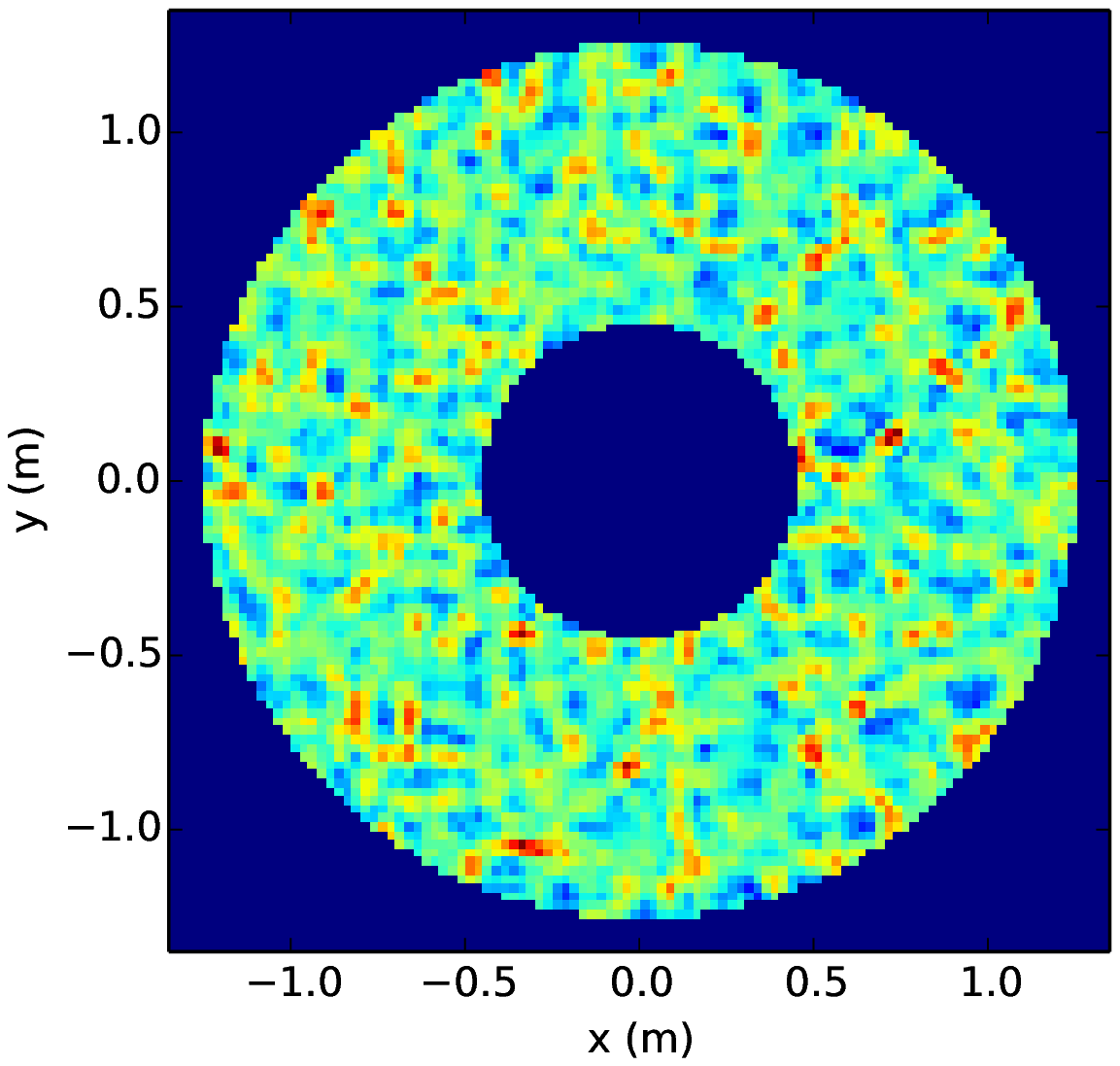} &
	\hspace{-5mm}\includegraphics[width=0.5\columnwidth]{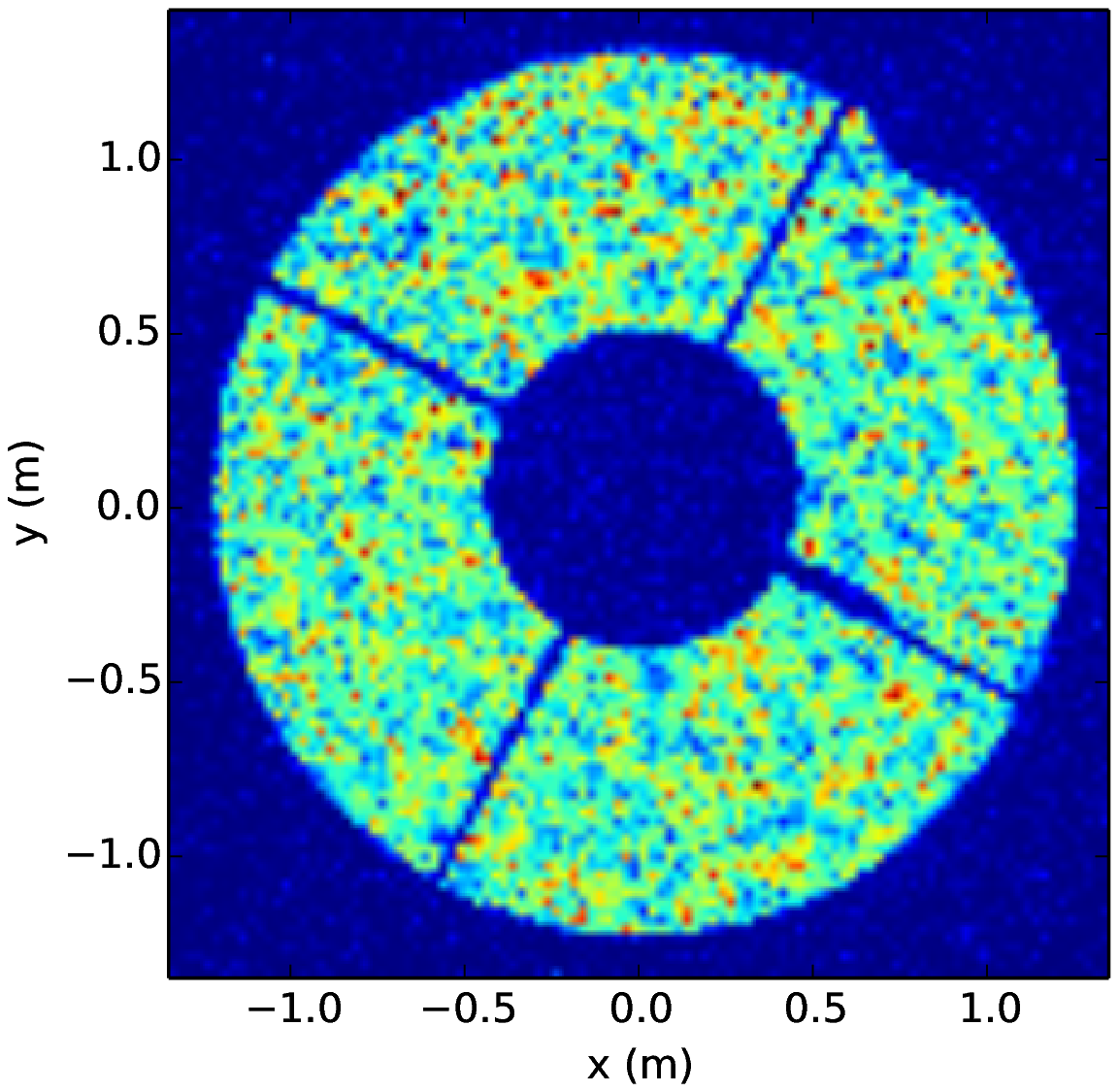}
\end{array}$
	\caption{Example simulated pupil image for a 2.5~m telescope and a single turbulent layer at 10~km (left) and an example of a real pupil image from the 2.5~m Isaac Newton Telescope on La Palma (right).}
	\label{fig:pupil_image}
\end{figure}

The effect of optical turbulence is twofold. The first effect is to deform the wavefront by retarding the sections passing through regions of higher refractive index. This limits the angular resolution of ground-based telescopes and is a first order effect as it depends on the first derivative (i.e. local tilt) of the wavefront. The second effect of the turbulence is to locally focus and defocus the wavefront, which results in spatial intensity fluctuations, or speckles, in the pupil plane of a telescope (figure~\ref{fig:pupil_image}). This is known as scintillation. Scintillation is a second order effect as it is caused by the second derivative or curvature of the wavefront. It is the propagation of this curvature that leads to the intensity speckles, therefore it is the higher altitude turbulence that is primarily responsible for this effect. This is different to the phase aberrations causing images to blur, which is dominated by the strongest turbulent layer, often near the ground \citep{Osborn10}. Therefore, atmospheric seeing (a measure of image quality) and photometric quality can be independent. It is possible to observe conditions which have bad seeing but good photometric quality.

The characteristic size of the scintillation speckles is given by the radius of the first Fresnel zone, $r_F = \sqrt{z\lambda}$, where $z$ is the propagation distance from the turbulent layer and $\lambda$ is the wavelength of the light. As a wavefront propagates away from a turbulent layer, increasing $z$ and hence $r_F$, the spatial intensity fluctuations become larger in terms of spatial extent. This is not dependent on the strength of the layer, which only affects the magnitude of the intensity fluctuations and not their spatial properties; for example, a turbulent layer at 10~km observed at 500~nm results in speckles of size $\sim$0.07~m, irrespective of the strength of the turbulence. 

The intensity speckles traverse the pupil with a velocity determined by the wind velocity of the turbulent layer. We assume that the speckle evolution timescale is longer than the wind crossing time. Speckles from different layers move independently and superimpose in the pupil plane. As the regions of higher intensity enter and exit the pupil the integrated total intensity also varies. The faster the wind speed the quicker the intensity will vary, demonstrating the importance of the wind speed in estimating the scintillation noise. It is these variations which lead to scintillation noise, which can limit the precision of photometric measurements. 

The amount of scintillation is referred to as the scintillation index, $\sigma_I^2$, and is expressed as the variance of the normalised intensity from the astronomical target. Assuming no other noise sources,
\begin{equation}
 \sigma_{I}^2=\frac{\left<I^2\right>-\left<I\right>^2}{\left<I\right>^2},
\end{equation}
where $I$ is the measured intensity from the object as a function of time and $\left<\right>$ denotes an ensemble average. The scintillation noise is therefore defined as $\sqrt{\sigma_{I}^2}$, and is the normalised scintillation noise.

\begin{figure}
	\centering
	\includegraphics[width=\columnwidth]{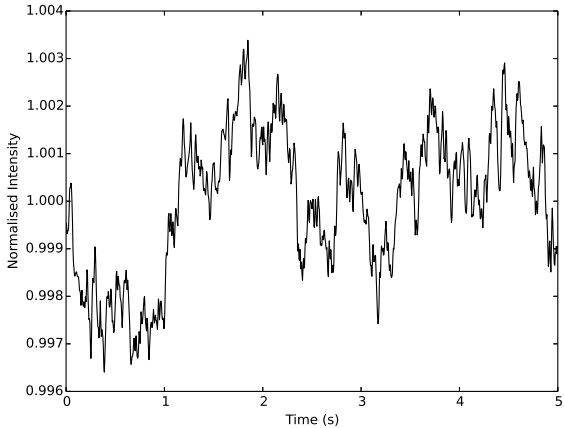}
	\caption{Example section of a simulated scintillation light curve. The intensity variations here are due entirely to atmospheric scintillation. In this case the telescope has a diameter of 8~m and the atmospheric turbulence is given by the mean profile for Paranal.}
	\label{fig:sim_int}
\end{figure}

\begin{figure}
	\centering
	\includegraphics[width=\columnwidth]{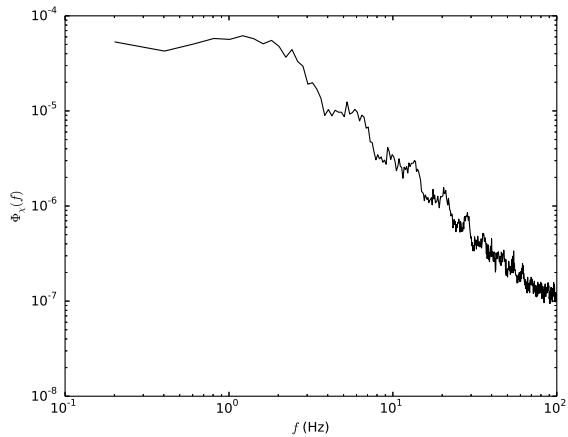}
	\caption{Scintillation power spectrum of the light curve shown in figure~\ref{fig:sim_int}. The power spectrum shows significant scintillation power over a wide range of frequencies.}
	\label{fig:sim_PS}
\end{figure}

Monte-Carlo computer simulations allow us to study the structure and effect of atmospheric scintillation independently of any other phenomena. Figure~\ref{fig:sim_int} shows a 5~s section of a simulated scintillation light curve from an 8~m telescope. In this light curve we see noise covering a wide range of temporal frequencies, $f$. Figure~\ref{fig:sim_PS} shows the power spectrum of the scintillation light curve. 
The power spectrum, $\Phi_\chi(f)$, shows that scintillation noise can be split into two regimes. Above a certain frequency, the power spectrum shows a power-law dependence of -11/3. Below this frequency the spectrum is flat. This turning point is determined by the amount of spatial averaging of the scintillation and is dependent on the telescope diameter and the wind velocity.


\section{Estimating scintillation noise}
\label{sect:scint_est}
\subsection{Estimating scintillation noise with Young's approximation}
\label{sect:Young}

A standard and often implemented equation for calculating the expected scintillation variance at any particular site \changed{is Young's approximation. This approximation, first suggested by \cite{Young1967}, is intended as an approximation and not a precise prediction. Young's approximation} is given by
\begin{equation}
\sigma_{Y}^2 = 10\times10^{-6}D^{-4/3}t^{-1}(\cos{\gamma})^{-3} \exp{\left(-2h_\mathrm{obs}/H\right)},
\label{eqn:Youngs}
\end{equation}
where $D$ is the diameter of the telescope, $t$ is the exposure time of the observation, $\gamma$ is the zenith distance, $h_\mathrm{obs}$ is the altitude of the observatory and $H$ the scale height of the atmospheric turbulence, which is generally accepted to be approximately 8000~m. All parameters are in standard SI units.

\subsection{Estimating scintillation noise with Atmospheric Turbulence Profiles}
\label{sect:prof}

\changed{Young's equation is an empirical approximation based on observations and is not intended to provide a precise prediction. Therefore, a} more precise option is to use concurrent measurements of the atmospheric turbulence to estimate the scintillation noise. 

As the scintillation noise is caused by the intensity speckles entering and leaving the telescope pupil there are three regimes of scintillation noise:
\begin{itemize}
\item{For short exposures the intensity speckles appear frozen in the pupil and no temporal averaging occurs.}
\item{For longer exposures the speckles traverse the pupil during an exposure and the scintillation noise will be reduced by temporal averaging. The amount by which the scintillation noise is reduced is dependent on the wind speed.}
\item{For small telescopes of approximately the same size as the speckles and smaller there will be a significant wavelength dependence. This is because the size of the speckles, $r_F$, is wavelength dependent.}
\end{itemize}

The scintillation index for short exposures is given by \cite{Kenyon06} (see appendix~\ref{sect:app} for details),
\begin{equation}
	\sigma_{I,\mathrm{se}}^{2} = 17.34D^{-7/3}\left(\cos\gamma\right)^{-3}\int_0^{\infty} h^{2}C_{n}^{2}\left(h\right)\mathrm{d}h,
\label{eq:scint_se}
\end{equation}
and for long exposures,
\begin{equation}
	\sigma_{I,\mathrm{le}}^{2} = 10.66D^{-4/3}t^{-1}\left(\cos\gamma\right)^{\alpha}\int_0^{\infty} \frac{ h^{2}C_{n}^{2}\left(h\right)}{V_{\bot}(h)} \mathrm{d}h,
	\label{eq:scint_le}
\end{equation}
where $C_n^2(h)$ is the profile of the refractive index structure parameter, a measure of the optical turbulence strength, $h$ is the altitude of the turbulent layer, with $h=z\cos{(\gamma)}$, $V_{\bot}(h)$ is the wind velocity profile and $\alpha$ is exponent of the airmass and is usually taken to be -3.5. Note that $\alpha$ will depend on the wind direction and vary between $\left(\cos\gamma\right)^{-3}$ for the case when the wind is transverse to the azimuthal angle of the star, and $\left(\cos\gamma\right)^{-4}$ in the case of a longitudinal wind direction. This difference comes from geometry; if the wind direction is parallel to the azimuth of the star, the projected pupil onto a horizontal layer is stretched by a factor of $1/\cos\gamma$.

We can re-arrange and solve equations~\ref{eq:scint_se} and \ref{eq:scint_le} for exposure time to find an expression for $t_\mathrm{knee}$, the exposure time at which the scintillation noise moves from the short exposure to long exposure regime,
\begin{equation}
t_\mathrm{knee} = 0.62 D (\cos\gamma)^{\alpha+3} \int_{0}^{\infty} \frac{1}{V_{\bot}(h)} dh,
\end{equation}
where $-1 \le \alpha+3 \le 0$ depending on the wind direction being parallel or perpendicular to the azimuthal angle of the target, respectively. This change from the short exposure regime to the long exposure regime can be seen in the scintillation power spectrum shown in figure~\ref{fig:sim_PS}. For atmospheric scintillation the change of regime occurs at an exposure time of approximately a few hundredths to a few tenths of a second.

It is interesting to note that in both the short-exposure and long-exposure regimes, scintillation noise is independent of wavelength. Therefore, in situations where the observations are scintillation dominated and one is observing with a multi-band instrument, e.g. ULTRACAM \citep{Dhillon07}, correcting one wavelength channel by another can reduce the scintillation noise, assuming that the variability that one wishes to measure is restricted to one band. \changed{However, although the scintillation noise is independent of wavelength, as observations move away from zenith the scintillation signals become temporally separated due to the chromatic dispersion in the atmosphere \citep{Dravins1997b}. This effect is exacerbated when the wind velocity is aligned with the azimuthal angle of the target. In addition,} if the observation is not scintillation dominated this action can be detrimental to the total photometric noise as the other noise sources will add in quadrature. 

For small apertures, where the aperture size is smaller than the spatial scale of the intensity fluctuations ($D<r_F$, i.e. $D$ is smaller than a few tens of centimetres) the scintillation index can be approximated using the equation given by \cite{Dravins1998},
\begin{equation}
	\sigma_{I}^{2} = 19.2\lambda^{-7/6}\left(\cos\gamma\right)^{-11/6}\int_0^{\infty} h^{5/6}C_{n}^{2}\left(h\right)\mathrm{d}h.
	\label{eq:scint_smallD}
\end{equation}

Several atmospheric optical turbulence profiling instruments are now in regular use at many of the world's premier observing sites. It is therefore possible to estimate the scintillation noise using techniques such as MASS (Multi Aperture Scintillation System, \citealp{Tokovinin07}), SCIDAR (SCIntillation Detection And Ranging, \citealp{Vernin73}) and SLODAR (SLOpe Detection And Ranging, \citealp{Wilson02}). However, only the SLODAR and SCIDAR methods can also measure the profile of the wind velocity as a function of altitude, as required by equation~\ref{eq:scint_le}.

\begin{figure*}
	\centering
	\includegraphics[width=2\columnwidth]{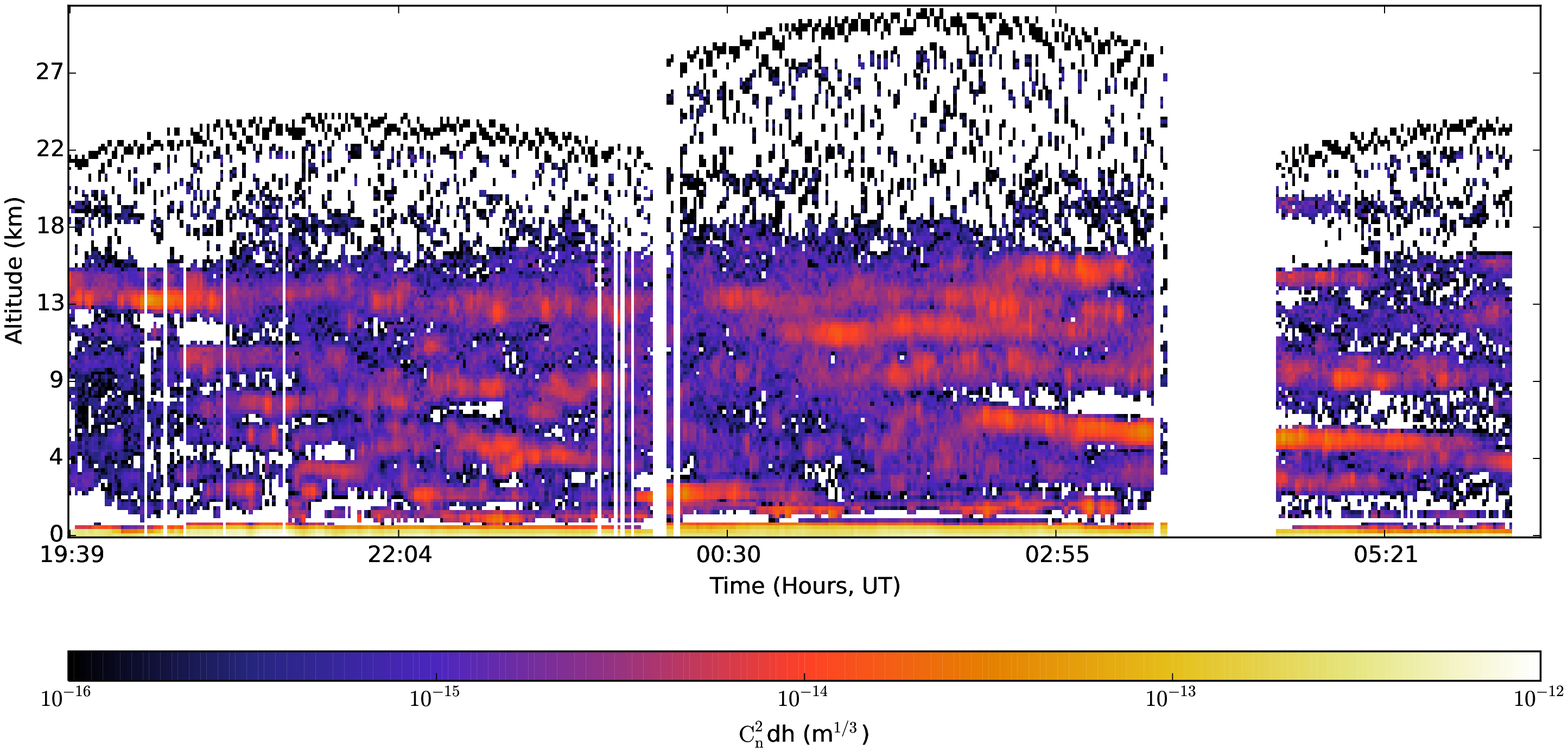}\\
	\vspace{-8mm}
	\includegraphics[width=2\columnwidth]{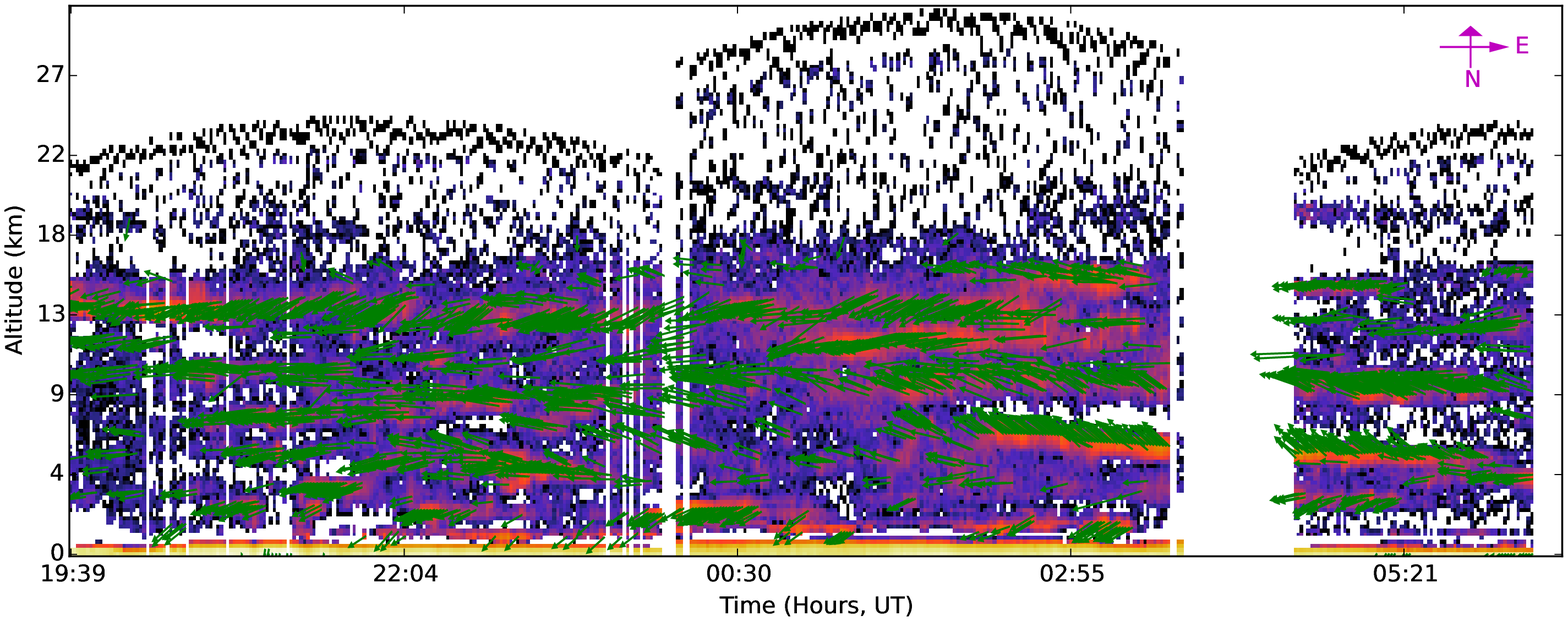}
	\caption{Atmospheric optical turbulence profiles for the night of 2014/10/09 from stereo-SCIDAR on the INT. The upper plot shows the turbulence strength profile and the lower plot shows the wind velocity profile. The colour-scale indicates the strength of the optical turbulence and the arrow length and direction indicate the turbulence speed and direction at a given time and altitude.}
	\label{fig:SS_profs}
\end{figure*}

Here we use stereo-SCIDAR \citep{Shepherd13, Osborn13}, a new high-resolution SCIDAR instrument, which can automatically obtain concurrent turbulence velocity and turbulence strength profiles. Stereo-SCIDAR was installed on the 2.5~m Isaac Newton Telescope (INT), La Palma, for three two-week campaigns in 2014 and on the 1~m Jacobus Kapteyn Telescope (JKT) for three two-week campaigns in 2013. An example atmospheric optical turbulence profile for a whole INT night is shown in figure~\ref{fig:SS_profs}. 

\subsection{Comparing measured scintillation noise with Young's approximation}
\begin{figure}
	\centering
	\includegraphics[width=\columnwidth]{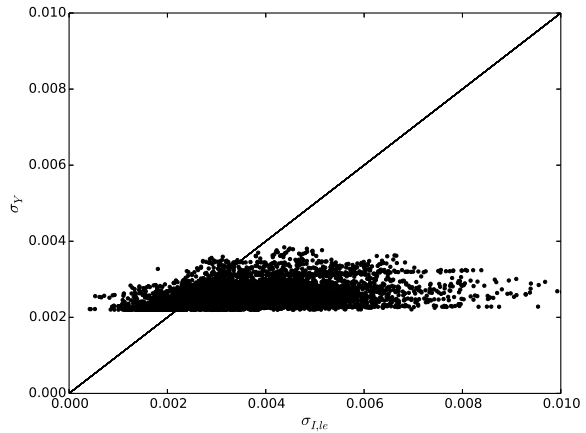}
	\caption{Scintillation noise comparison between Youngs's approximation (equation~\ref{eqn:Youngs}) and estimated scintillation noise from real SCIDAR profiles (equation~\ref{eq:scint_le}). We see that Young's approximation tends to underestimate the scintillation noise as measured by SCIDAR. We also see that SCIDAR measures a much greater range of scintillation noise values. This is due to the variability of the atmospheric turbulence profile. The solid line represents $\sigma_Y = \sigma_{I,\mathrm{le}}$, and is where the points should lie if equations \ref{eqn:Youngs} and \ref{eq:scint_le} agree.}
	\label{fig:scint_young}
\end{figure}

Figure~\ref{fig:scint_young} demonstrates the difference between the long exposure scintillation noise as estimated by Young's approximation (equation~\ref{eqn:Youngs}) and that calculated from observed SCIDAR profiles (equation~\ref{eq:scint_le}). The scatter in Young's approximation of the scintillation noise is purely due to varying airmass during the observations. The scatter in the SCIDAR scintillation noise is due to the varying airmass, the contribution of the wind direction component in the exponent of the airmass, and the inherent atmospheric variability; such variability is not included in Young's approximation.

\begin{figure}
	\centering
	\includegraphics[width=\columnwidth]{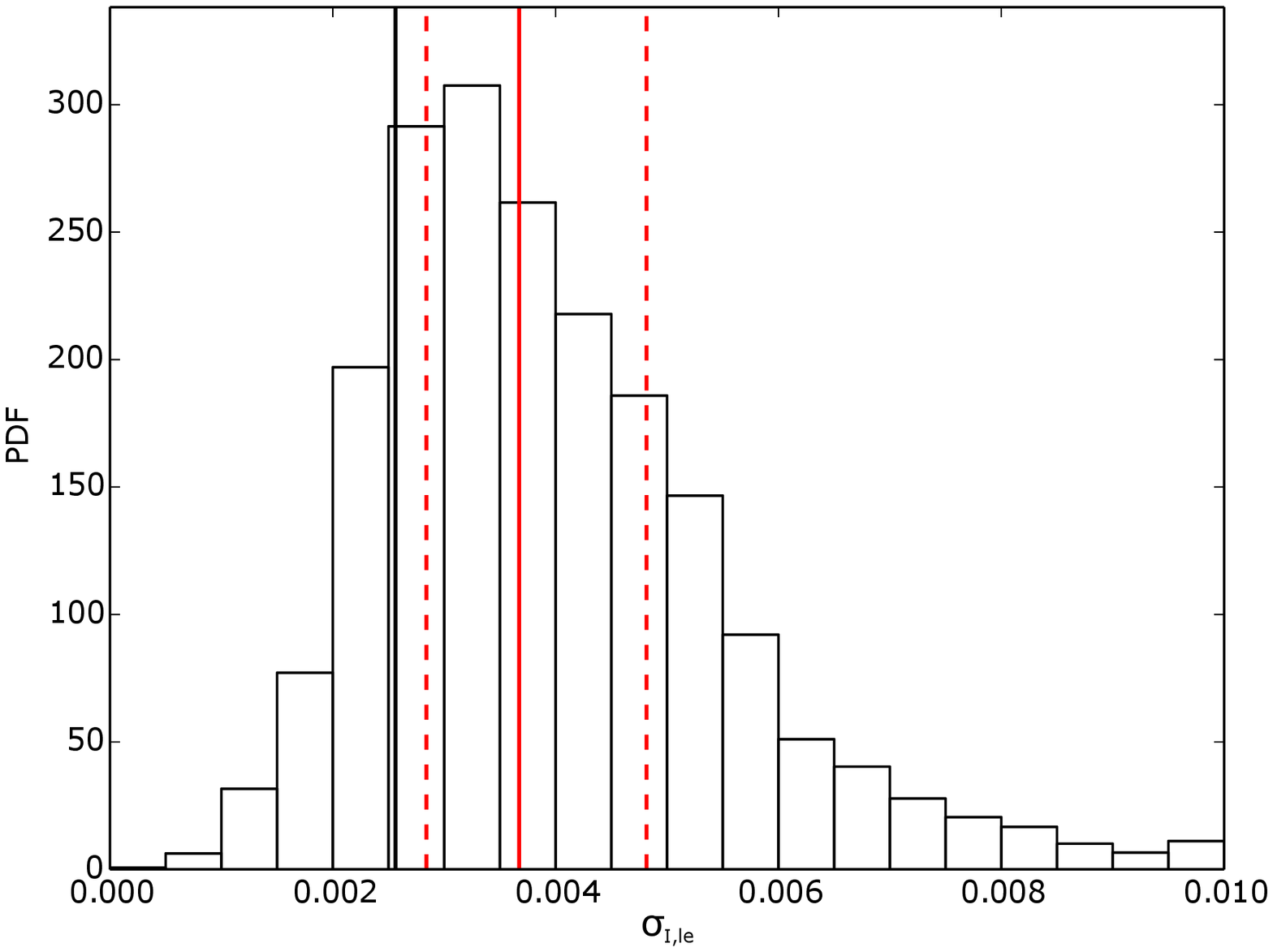}
	\caption{Probability Density Function (PDF) of scintillation noise as estimated from real SCIDAR profiles (equation~\ref{eq:scint_le}) for a 1~m telescope and 1~s exposure times. The solid and dashed red lines indicate the median and 1st and 3rd quartiles, respectively. The solid black line indicates the value given by Young's approximation (equation~\ref{eqn:Youngs}) at the zenith.}
	\label{fig:scint_dist}
\end{figure}

Figure~\ref{fig:scint_dist} shows the distribution of the scintillation noise predicted from the SCIDAR profiles using equation~\ref{eq:scint_le}. The assumed telescope size was 1~m and the exposure time was 1~s. The normalised scintillation noise has a log-normal distribution with median 0.0036 and 1st and 3rd quartiles of 0.0028 and 0.0048 respectively. For comparison, the scintillation noise at zenith as estimated by Young's equation is 0.0026. Clearly, there is a large variability in observed scintillation noise which is not reflected in Young's single-value approximation (equation~\ref{eqn:Youngs}).

\begin{figure}
	\centering
	\includegraphics[width=\columnwidth]{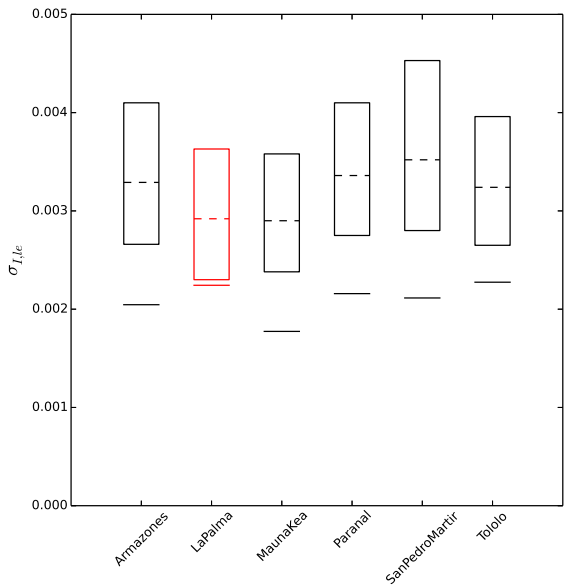}
	\caption{Expected and measured scintillation noise for major observatories around the world. The bottom and top of the boxes indicate the 1st and 3rd quartiles and the dashed line shows the median measured scintillation noise. The solid line external to the box indicates the expected value given by Young's approximation (equation~\ref{eqn:Youngs}). The data for La Palma is calculated from recent results from Stereo-SCIDAR on the INT, and the others are extracted from \protect\cite{Kornilov12d}.}
	\label{fig:scint_box}
\end{figure}

In figure~\ref{fig:scint_box} we compare measured scintillation statistics from several major observatories around the world \citep{Kornilov12d} including the values we have determined for La Palma in figure~\ref{fig:scint_dist}. We show the median, 1st and 3rd quartile scintillation noise and the calculated Young's estimate for each site. From this plot we see that all of the observatories in this study show similar scintillation statistics. However, La Palma and Mauna Kea have the lowest median scintillation noise. San Pedro de M\'artir (Mexico), Armazones and Paranal have the highest median scintillation noise. Mauna Kea has the smallest interquartile range suggesting stable low-scintillation conditions, whilst San Pedro de M\'artir has the largest interquartile range, demonstrating that the conditions can be more variable there. \changed{However, it should be noted that the data shown here is for a limited set of data and as we have shown scintillation noise can be extremely variable. In addition, we are only examining the variance of the intensity fluctuations and have not included other parameters such as the characteristic timescale, which will have a strong dependence on the typical wind velocity profile and may be important for high-speed photometry.}

Figure~\ref{fig:scint_box} confirms that Young's approximation underestimates the measured scintillation noise at all sites, typically by a factor of 1.5. During any one night the scintillation noise will usually not be equal to the median value and a large variation in scintillation noise can be observed, even at the best locations. Figure~\ref{fig:scint_box} also shows that the best astronomical sites for seeing are not necessarily the best for scintillation. For example, Mauna Kea and Paranal have similar seeing statistics (for example \citealp{Sarazin08,Chun09}) but the former is better than the latter for scintillation noise according to the data shown in figure~\ref{fig:scint_box}. One explanation for this could be that the high-altitude turbulence is weaker and more stable at Mauna Kea.

It is possible to add a coefficient to Young's approximation to make it match the measured median scintillation. The modified Young's approximation becomes
\begin{equation}
\sigma_{Y}^2 = 10\times10^{-6}C_Y^2 D^{-4/3}t^{-1}(\cos{\gamma})^{-3} \exp{\left(-2h_\mathrm{obs}/H\right)},
\label{eqn:Youngs2}
\end{equation}
where $C_Y$ is the empirical coefficient listed in table~\ref{tab:Young_Coeff} for several major observatories (with a mean value of $1.5$). The coefficients for the 1st and 3rd quartiles are also listed to enable the variability of the expected scintillation noise at each location to be estimated.
\begin{table}
\caption{Values of the empirical coefficient, $C_Y$, in our modified Young's approximation (equation~\ref{eqn:Youngs2}) at a selection of observatories for the median ($\mathrm{C}_\mathrm{median}$), 1st quartile ($\mathrm{C}_\mathrm{Q1}$) and 3rd quartile ($\mathrm{C}_\mathrm{Q3}$) of measured scintillation noise.}
\label{tab:Young_Coeff}
\begin{tabular}{@{}lccc}
\hline
Observatory & $\mathrm{C}_\mathrm{median}$ & $\mathrm{C}_\mathrm{Q1}$ & $\mathrm{C}_\mathrm{Q3}$\\
\hline
Armazones & 1.61 & 1.30 & 2.00\\
La Palma & 1.30 & 1.02 & 1.62\\
Mauna Kea & 1.63 & 1.34 & 2.02\\
Paranal & 1.56 & 1.27 & 1.90\\
San Pedro de M\'artir & 1.67 & 1.32 & 2.14\\
Tololo & 1.42 & 1.17 & 1.74\\
\hline
\end{tabular}
\end{table}

We recommend that astronomers use our modified form of Young's approximation (equation~\ref{eqn:Youngs2}) with the $C_Y$ values listed in table~\ref{tab:Young_Coeff} to give more reliable estimates of the scintillation noise. For even more precise estimates, or if the observing site is not listed in table~\ref{tab:Young_Coeff}, it is best to use median atmospheric turbulence profiles (which exist for most observatories) in equation~\ref{eq:scint_le}. 

\subsection{Real-time scintillation estimation}

Real profiles from stereo--SCIDAR show that the optical turbulence profile (both turbulence strength and velocity) can evolve rapidly. We have seen periods of strong, high layers which survive for several hours and we have also seen small bursts of activity lasting only a few minutes \citep{Shepherd13, Avila98}. The altitude and strength of these layers varies with time, having significant impact on the scintillation noise during observations. This variability in measured scintillation noise can be seen in figures~\ref{fig:SS_profs} -- \ref{fig:scint_box}.

Access to concurrent turbulence profiles allows the real-time estimation of scintillation noise using equations~\ref{eq:scint_se}, \ref{eq:scint_le} and \ref{eq:scint_smallD}. As the scintillation noise is dominated by high-altitude turbulence it is likely to be isotropic and consistent over the entire site. This can be validated by the fact that we see the same high-altitude turbulence profile before and after we change target with the profiler, with one target setting in the west as the other rises in the east. This is not true with low-altitude turbulence which is dependent on the local topography, including the telescope dome and structure.

\begin{figure}
	\centering
	\includegraphics[width=\columnwidth]{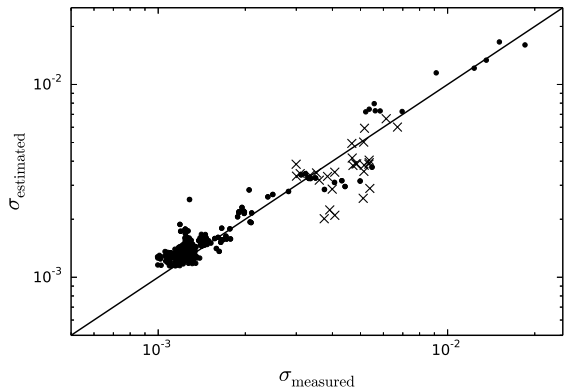}
	\caption{Comparison of measured noise, $\sigma_{\mathrm{measured}}$, with estimated noise, $\sigma_\mathrm{estimated}$ for photometric data, including scintillation, shot and detector noise. The measured noise values were obtained from light curves observed with the 4.2~m WHT (points) and a 0.5~m telescope on the same site (crosses), using exposure times between 5 and 10~s. The estimated noise values were obtained from SCIDAR profiles. The correlation coefficient is 0.96.}
	\label{fig:scint_comp}
\end{figure}

Figure~\ref{fig:scint_comp} shows a comparison of the noise measured from light curves with the noise estimated from SCIDAR profiles obtained at the same site and time, including the effects of shot and detectors noise. In most cases scintillation was seen to be a significant source on noise. See \cite{FohringThesis14} for details. The large correlation coefficient of 0.96 implies that concurrent atmospheric optical turbulence profiles from an external profiler can be used to precisely model the noise in a photometric light curve. This enables the scintillation noise of an observation to be estimated, which can then be used for instrument characterisation, data validation and theoretical model fitting.

\section{Scintillation noise in Photometric measurements}
\label{sect:meas}

\begin{figure}
	\centering
	\includegraphics[width=\columnwidth]{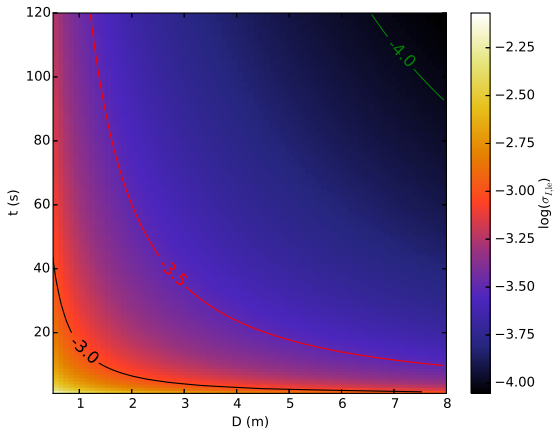}
	\caption{Theoretical long exposure scintillation noise as a function of exposure time and telescope diameter. The scintillation noise was calculated for median atmospheric conditions on La Palma and varies between 1\% for small telescopes and short exposure times, and 0.01\% for larger telescopes and longer exposure times.}
	\label{fig:theory_scint}
\end{figure}

Figure~\ref{fig:theory_scint} shows the expected long exposure scintillation noise for varying exposure times and telescope diameters. The scintillation noise was calculated for median atmospheric conditions on La Palma from equation~\ref{eq:scint_le} and varies between 1\% for small telescopes and short exposure times, and 0.01\% for larger telescopes and longer exposure times. 

The fractional shot noise ($\sqrt{S}/S$, where $S$ is the signal) can be estimated using
\begin{equation}
\sigma_{s} = 1/\sqrt{At\eta\delta\lambda\phi},
\label{eq:shot}
\end{equation}
where $A$ is the collecting area of the telescope, $\eta$ is the throughput (including optics and CCD quantum efficiency), $\delta\lambda$ is the bandwidth and $\phi$ is the flux of photons incident on the telescope. 

When observing bright targets the sky background and readout noise can be considered negligible in comparison with the shot noise and scintillation noise. Therefore, we only consider the latter two noise terms. The scintillation noise is given by, $\sigma_{I} = \sqrt{\sigma_{T}^2 - \sigma_{s}^2}$, where $\sigma_T$ is the total photometric noise. The fractional shot noise is given by $\sigma_{s} = 1/\sqrt{\left<S\right>}$, where $\left<S\right>$ is the time-averaged signal, where the scintillation noise is expected to approach zero.

\begin{figure}
	\centering
	\includegraphics[width=\columnwidth]{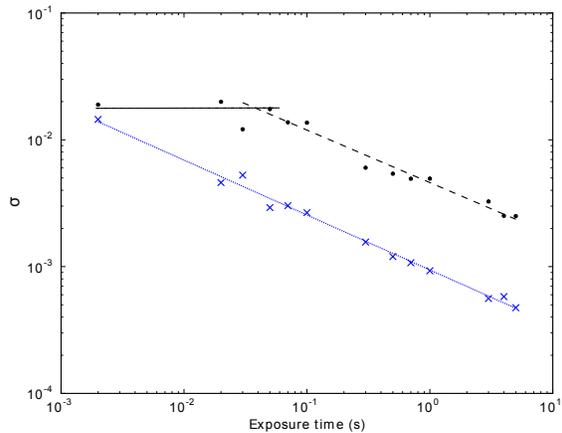}
	\caption{Photometric noise as a function of exposure time on the JKT, for a magnitude 8 star observed on 21/07/2013. The points indicate the estimated scintillation noise, $\sigma_I$, and the crosses are the estimated fractional shot noise, $\sigma_s$. The lines are the theoretical counterparts to each segment given by equations~\ref{eq:scint_se} (short exposure; dot-dashed line), \ref{eq:scint_le} (long exposure; dashed line) and \ref{eq:shot} (fractional shot noise; dotted line).}
	\label{fig:measured_scint}
\end{figure}

Figure~\ref{fig:measured_scint} shows the estimated shot and scintillation noise in the short- and long-exposure regimes from on-sky data obtained with the 1~m Jacobus Kapteyn Telescope (JKT) on La Palma. As the exposure time is increased, the scintillation noise remains constant until we reach $t_\mathrm{knee}\sim0.05$~s; at this point the scintillation changes from the short-exposure regime (with no temporal averaging), described by equation~\ref{eq:scint_se}, to the long-exposure regime (equation~\ref{eq:scint_le}). For exposures longer than $t_\mathrm{knee}$ the scintillation noise reduces with a $1/\sqrt{t}$ dependence. It is interesting to note that in the long-exposure regime, the fractional shot noise has the same time dependence as scintillation noise. Therefore, the ratio of scintillation noise to shot noise is conserved for all exposure times. The significance of this is that although longer exposure times will reduce the total noise, the ratio of the shot noise to the scintillation noise will remain fixed. 

\begin{figure*}
\centering
$\begin{array}{cc}
	 \includegraphics[width=0.5\textwidth]{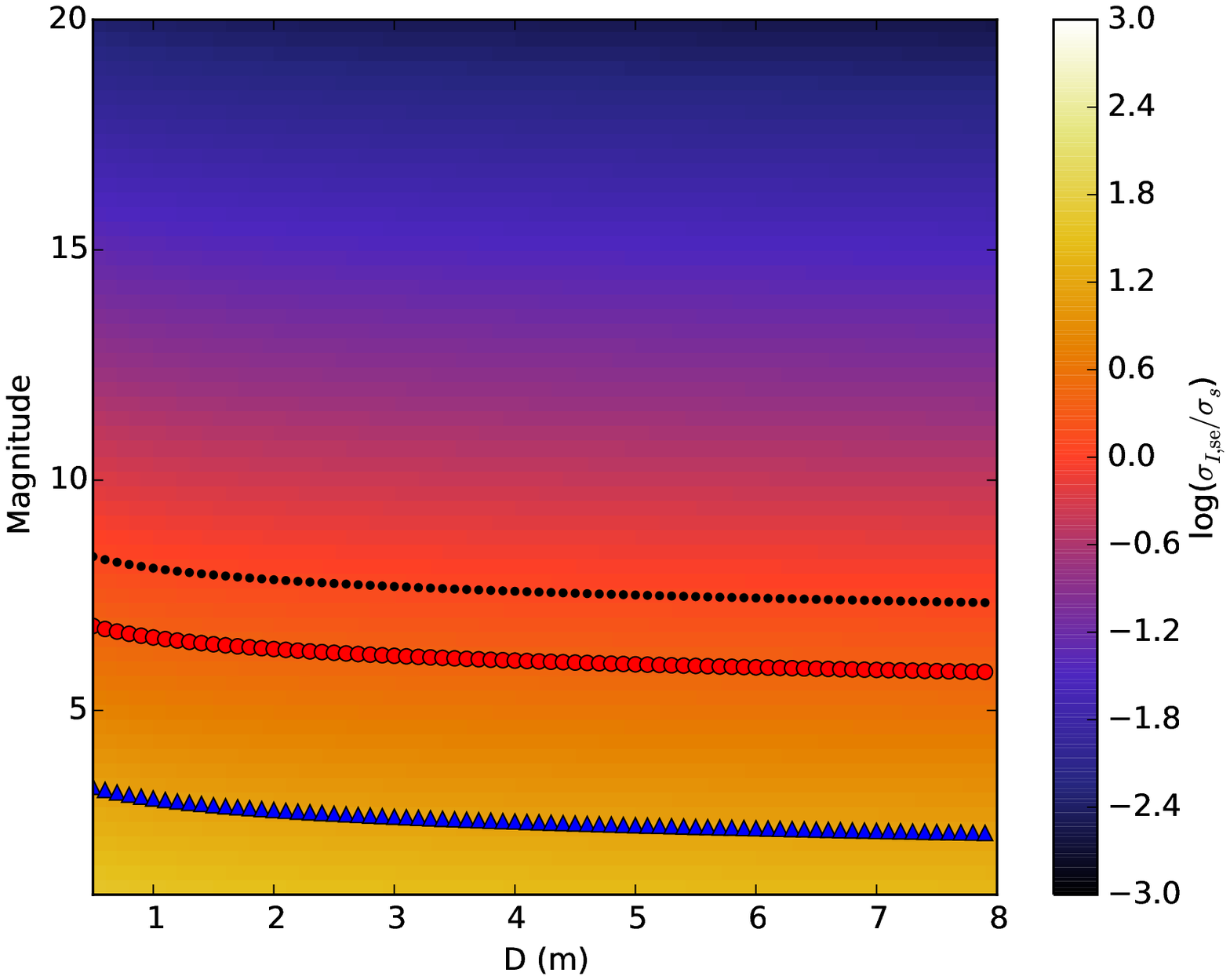}&
	 \includegraphics[width=0.5\textwidth]{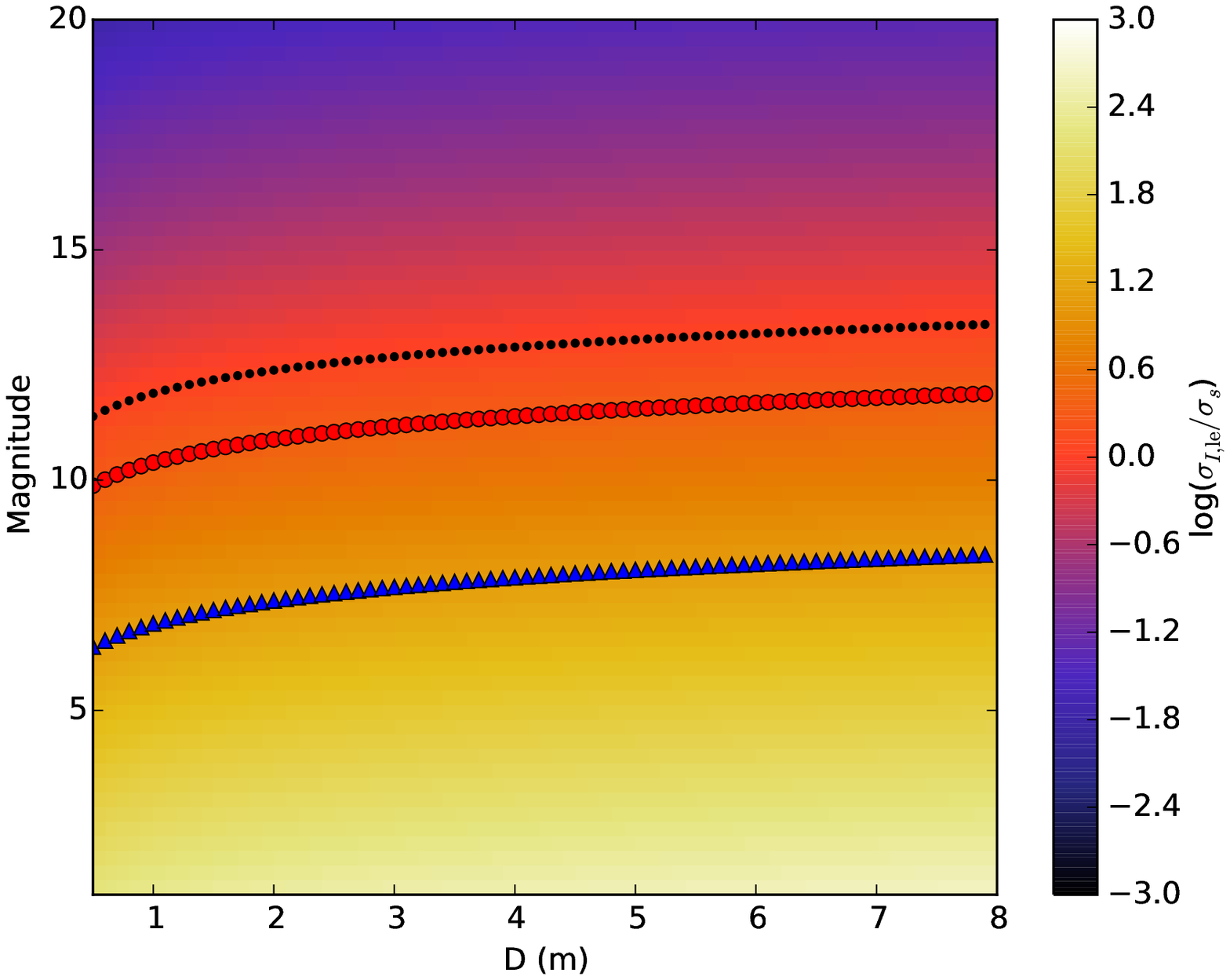}
\end{array}$
\caption{Theoretical parameter space plots for the ratio of the scintillation to shot noise in the short exposure regime (left) and the long exposure regime (right), for varying telescope diameter and target stellar magnitude ($V$-band). The short exposure time is set to 2~ms. The long exposure time is irrelevant as both noise sources have the same exposure time dependence, making the ratio independent of exposure time. The black dotted line shows where the scintillation noise equals the shot noise. For any telescope diameter / target magnitude combinations below this line, the scintillation noise is greater than the shot noise and vice versa. The red line composed of circles indicates a ratio of 2, i.e. when the scintillation noise is twice the shot noise. The blue line composed of triangles indicates the point where the scintillation noise is an order of magnitude larger than the shot noise.}
\label{fig:2Dscint_shot_comp}
\end{figure*}

Figure~\ref{fig:2Dscint_shot_comp} shows a 2D plot of the theoretical parameter space for the ratio of the scintillation noise to the shot noise for varying telescope diameters and object magnitudes. We see that when observing bright objects the scintillation noise dominates over the shot noise, and hence dominates the total photometric noise.

By combining equation~\ref{eq:shot} with equation~\ref{eq:scint_se} or \ref{eq:scint_le} we see that the ratio of scintillation noise to photon noise scales with telescope diameter as $\sigma_{I,\mathrm{se}}/\sigma_{s}\propto D^{-1/6}$ and  $\sigma_{I,\mathrm{le}}/\sigma_{s}\propto D^{1/3}$ for short and long exposures respectively. This is a weak dependence and so we can say that for median conditions and regardless of telescope diameter, scintillation will be greater than shot noise for ($V$-band) magnitudes less than $\sim$13 for long exposures and $\sim$8 for short exposures. It will dominate at magnitudes greater than $\sim$12 for long exposures and 6.5 for short exposures. These results are for instruments with 100\% throughput and at a zenith distance of 30~degrees; for other values, the results must be scaled accordingly. 

\section{Scintillation on large and extremely large telescopes}
\label{sect:largeTel}

Kolmogorov's model of turbulence used for the preceding equations assumes that energy is
injected into the turbulent medium at large spatial scales (the outer scale, $L_0$)
and forms eddies. These then break down into smaller eddies in a self-similar
cascade until the eddies become small enough that the energy is dissipated by
the viscous properties of the medium. This will occur at the inner scale, $l_0$. In the inertial range between the inner and outer scales Kolmogorov
predicted a turbulent phase power spectrum with distribution, $f^{-11/3}$. 

As scintillation is an effect of the curvature of the wavefront it has power on all spatial scales up to the outer scale of the turbulence. On smaller telescopes the effect of the outer scale is negligible as we can assume the outer scale to be considerably larger than the telescope diameter. However, as telescope sizes approach that of the outer scale (of the order 10 -- 100~m) it should be included in theoretical calculations. The inner scale is generally neglected as it is accepted to be a few millimetres in the atmospheric case and too small to have any significant impact.

All of the equations above (\ref{eqn:Youngs}, \ref{eq:scint_se}, \ref{eq:scint_le}, \ref{eq:scint_smallD} and \ref{eqn:Youngs2}) assume infinite outer scale, i.e. Kolmogorov statistics. The next generation of extremely large telescopes of the scale 20 to 40~m are currently under construction. These telescopes will be of the same spatial scale, and potentially even larger, than the outer scale of the turbulence. If we want to examine the dynamics of systems with spatial scales approaching the outer scale then we need to modify the equations that we use.  

The scintillation index for extremely large telescopes is expressed as (see appendix~\ref{sect:app} for derivation)
\begin{multline}
\sigma_{I,\mathrm{le}}^2 = 12.24D^{-4/3} t^{-1}  \left(\cos\gamma\right)^{\alpha} \int \frac{C_n^2(h) h^{2}}{V_\bot(h)}  \\
\int \left(q^2 + (D/L_0(h)^{2})\right)^{-11/6} q^2 (\mathrm{J}_1(\pi q))^2 \mathrm{d}q \mathrm{d}h,
\label{eq:scint_gen}
\end{multline}
where $q=Df$, and can be solved numerically for the particular telescope and the prevailing atmospheric turbulence conditions. 

\begin{figure}
	\centering
	\includegraphics[width=\columnwidth]{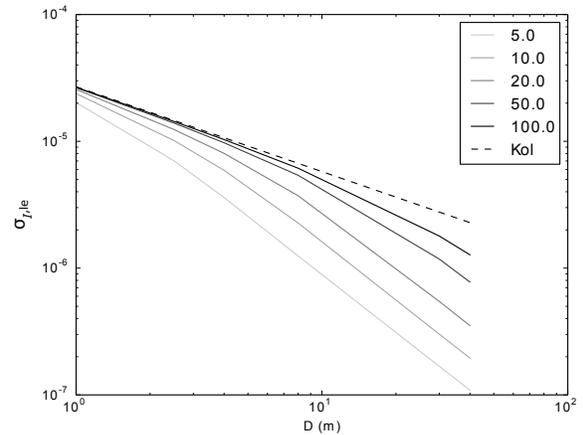}
	\caption{Theoretical scintillation noise for telescope diameters between 1~m and 40~m for the median atmospheric turbulence profile on La Palma in the long exposure regime ($t=1$~s). The dashed line is the theoretical value calculated assuming an infinite outer scale (i.e. Kolmogorov turbulence; equation~\ref{eq:scint_le}). The solid lines indicate finite outer scales with values of 100, 50, 20, 10, and 5~m. An outer scale of 10~m results in the scintillation noise being up to an order of magnitude less than the value given by equation~\ref{eq:scint_le} for a 39~m telescope, such as the European Extremely Large Telescope.}
	\label{fig:L0_scint_long}
\end{figure}

Figure~\ref{fig:L0_scint_long} shows the expected scintillation noise for median conditions on La Palma for telescope sizes between 1~m and 40~m in the long exposure regime. The expected scintillation noise is always lower than that predicted for infinite outer scale, i.e. Kolmogorov turbulence (equation~\ref{eq:scint_le}), and this difference increases with increasing diameter.

The factor by which equation~\ref{eq:scint_le} overestimates the scintillation noise is dependent on $L_0$. \changed{The outer scale of turbulence is currently poorly understood and estimates of its value are uncertain and varied. Therefore the effect of outer scale on the scintillation noise on large telescopes is difficult to estimate, but }it can be seen that, even for existing large telescopes of 8-10~m  diameter, the outer scale will reduce the expected scintillation noise. \changed{The next generation of extremely large telescopes will all have segmented pupils which will complicate the situation as the gaps between the segments will increase the high frequency scintillation components in a non-trivial way \citep{Dravins1998}. In order to estimate this effect the aperture filter function, $A(f)$, will need to be defined for each case and implemented into the equations above. The aperture filter function of an arbitrary pupil function, $p(x,y)$ defined in a cartesian coordinate system of $x$ and $y$ basis, is,
\begin{equation}
A(f) = \mid \mathcal{F} \left(P(x,y)\right) \mid^2.
\end{equation}
Scintillation noise estimation on segmented or non-circular telescopes will therefore need to be examined on a case by case basis and can be done by replacing the bespoke aperture filter function into the derivation in appendix~\ref{sect:app}.} Scintillation noise on large telescope could be overestimated by an order of magnitude \citep{Kornilov12b}. This suggests that extremely large telescopes will be even more of a valuable facility for high-precision photometry than previously thought.

\section{Scintillation noise with central obscurations}
\label{sect:secObs}
It is also important to study the impact of the central obscuration of the telescope on scintillation noise. Larger secondary mirrors lead to more scintillation noise due to the smaller collecting area over which the scintillation speckles are spatially averaged. 

The scintillation index as a function of the telescope central obscuration in the short-exposure regime is given by (see appendix~\ref{sect:app})
\begin{multline}
\sigma_{I,\mathrm{se}}^2 = 38.44D^{-7/3}  \left(\cos\gamma\right)^{-3} \int C_n^2(h) h^{2}  \\
\int \left(q^2 + (D/L_0(h)^{2})\right)^{-11/6} q^3 \\
(\mathrm{J}_1(\pi q) - \epsilon \mathrm{J}_1(\pi \epsilon q))^2 / (1-\epsilon^2)^2 \mathrm{d}q \mathrm{d}h,
\label{eq:scint_gen_short}
\end{multline}
and in the long-exposure regime is given by,
\begin{multline}
\sigma_{I,\mathrm{le}}^2 = 12.24D^{-4/3} t^{-1}  \left(\cos\gamma\right)^{\alpha} \int \frac{C_n^2(h) h^{2}}{V_\bot(h)}  \\
\int \left(q^2 + (D/L_0(h)^{2})\right)^{-11/6} q^2 \\
(\mathrm{J}_1(\pi q) - \epsilon \mathrm{J}_1(\pi \epsilon q))^2 / (1-\epsilon^2)^2 \mathrm{d}q \mathrm{d}h,
\label{eq:scint_gen_long}
\end{multline}
where $\epsilon$ is the ratio of the diameter of the central obscuration to the diameter of the primary mirror.

Using the above equations it is possible to calculate the scintillation noise for any telescope diameter and in any atmospheric conditions. However, for telescopes smaller than approximately 5~m the outer scale of the turbulence can be ignored and equations~\ref{eq:scint_se}, \ref{eq:scint_le} and \ref{eq:scint_smallD} can be safely used. Most modern telescopes have a central obscuration of approximately 30\% in terms of diameter, or 10\% in area. In the long exposure regime this makes negligible difference to the scintillation noise. In the short exposure regime we would expect the scintillation noise to be increased by a factor of approximately 1.2. 

\changed{In the case of a large telescope, this increase in the scintillation noise due to the central obscuration needs to be compared to the expected reduction in scintillation noise due to outer scale effects (as discussed in section~\ref{sect:largeTel}) \citep{Kornilov12c}.}

\section{Discussion}

Below we summarise the key points about scintillation noise that astronomers should remember when performing astronomical photometry:
\begin{itemize}
\item{
Scintillation is dominated by high-altitude turbulence, with very little scintillation being caused by low-altitude turbulence. This is different to atmospheric seeing which blurs astronomical images. Seeing is dominated by the strongest layers of atmospheric turbulence, irrespective of altitude. In fact it is the surface layer which often dominates the seeing \citep{Osborn10}. Therefore, it is possible for the seeing to be bad whilst the scintillation noise is low and vice versa.
}
\item{
The ratio of scintillation noise to shot noise is only weakly dependent on telescope diameter ($D^{-1/6}$ in the short exposure regime and $D^{1/3}$ in the long-exposure regime) and independent of exposure time. Therefore increasing the telescope diameter and exposure time makes little difference to the ratio of shot noise to scintillation noise.
}
\item{
Young's approximation (equation~\ref{eqn:Youngs}) is often used by astronomers to estimate the scintillation noise in their light curves. This equation, which includes the telescope diameter, airmass, exposure time and altitude above sea level, is only an approximation to the median scintillation noise \changed{and is not intended to be a precise estimate}. From recent Stereo-SCIDAR atmospheric profiles on La Palma and published data from \cite{Kornilov12d} we have found that Young's approximation consistently underestimates the fraction of scintillation noise by a mean value of 1.5 at a range of astronomical sites. 
}
\item{We have presented a modified form of Young's approximation (equation~\ref{eqn:Youngs2}) that uses empirical correction coefficients listed in table~\ref{tab:Young_Coeff} to give more reliable estimates of the scintillation noise at a range of astronomical sites.
}
\item{
If even greater precision is required than provided by our modified form of Young's approximation, it is possible to use the median atmospheric profile for an observatory in conjunction with equations~\ref{eq:scint_se} and \ref{eq:scint_le}. Of course the instantaneous turbulent profile is unlikely to match the median atmospheric turbulence profile, and hence for the ultimate in precision it is necessary to use contemporaneous atmospheric turbulence profiles with these equations.
}
\item{
The vertical profile of optical turbulence is very variable, even at the World's premier observing locations. Figure~\ref{fig:SS_profs} demonstrates this for one night at the Observatorio del Roque de los Muchachos, La Palma. Therefore, the median profile can not give a precise representation of the scintillation noise at any particular time. Here we propose to use atmospheric turbulence profiling instrumentation to provide a better estimate of the scintillation noise in high-precision photometric observations. Although atmospheric turbulence profilers remain uncommon, with the proliferation of adaptive optics (AO) systems they will become more widely available. Indeed it is possible to derive the required information from the AO system itself \citep{Cortes12}. 
}
\item{
A more precise knowledge of scintillation noise is important to enable performance assessment, calibration and optimisation of photometric instrumentation. It is also useful when fitting models to photometric data (for example, extrasolar planet eclipse light curves; \citealp{FohringThesis14}), and to help develop scintillation correction concepts.
}
\item{
As telescope diameters approach the outer scale of the optical turbulence, the measured scintillation noise will be lower than that predicted for infinite outer scale by Kolmogorov equations~\ref{eqn:Youngs}, \ref{eq:scint_se}, \ref{eq:scint_le}, \ref{eq:scint_smallD} and \ref{eqn:Youngs2}. This means that very large and extremely large telescopes ($\sim$8 - 40~m) are favourable facilities for high-precision photometry. We present modified equations (\ref{eq:scint_gen_short} and \ref{eq:scint_gen_long}) which should be used to calculate the scintillation noise with large telescopes or telescopes with a large secondary obscuration.
}
\item{There are several proposed scintillation mitigation techniques at varying stages of development, such as Conjugate-Plane Photometry \citep{Osborn11}, Tomographic wavefront reconstruction \citep{Osborn15} and active deformable mirror techniques \citep{Viotto12}. There are also passive techniques such as `lucky photometry', where only data taken during times of low scintillation noise is used, or wavelength correction (where one wavelength channel can be used to correct another; \citealp{Kornilov11b}). \changed{However, although the scintillation noise is independent of wavelength as observations move away from zenith the scintillation signals become temporally separated due to the chromatic dispersion in the atmosphere \citep{Dravins1997b}. This effect is exacerbated when the wind velocity is aligned with the azimuthal angle of the target. In addition,} wavelength correction will only work in situations where the scintillation noise definitely dominates. Otherwise the noise will be made worse by the correction process as independent noise sources (i.e. shot noise) will add in quadrature. It will also only work in the large telescope ($D\ge\sim0.1$~m) regime when the scintillation noise is almost independent of wavelength and it will not work if the variability that one wants to measure is present at both wavelengths. 
}
\end{itemize}

\section*{Acknowledgments}
We are grateful to the Science and Technology Facilities Committee (STFC) for financial support (grant reference ST/J001236/1). FP7/2013-2016: The research leading to these results has received funding from the European Community's Seventh Framework Programme (FP7/2013-2016) under grant agreement number 312430 (OPTICON). The William Herschel Telescope, the Isaac Newton Telescope and the Jacobus Kapteyn Telescope are operated on the island of La Palma by the Isaac Newton Group in the Spanish Observatorio del Roque de los Muchachos of the Instituto de Astrof'sica de Canarias.

\bibliographystyle{mn2e}
\bibliography{/Users/jo/Documents/Papers/mybibliography}

\appendix

\section{Derivation of scintillation equations}
\label{sect:app}

We can derive the theoretical scintillation index as the integral of the scintillation power spectrum \citep{Roddier1981}, 
\begin{equation}
\sigma_{I}^{2} = \int_0^\infty W(f) df
\label{eqn:scint_sigma_A}
\end{equation}
where $W(f)$ is the irradiance power spectrum, given by \citep{Tokovinin02},
\begin{equation}
W(f) = 9.7 \times 10^{-3} \times 4 \times (2\pi)^3 \int_0^{\infty} C_n^2(z) \phi(f) S(z,f) A(f) f \mathrm{d}z.
\label{eqn:scint_weight_A}
\end{equation}
$S(z,f)$ is the Fresnel filter function to account for the wavefront propagation and is given by $\sin^2\left(\pi\lambda z f^2\right)/\lambda^2$ \citep{Roddier1981}. It is this function that gives the intensity fluctuations an intrinsic spatial scale of $r_F=\sqrt{\lambda z}$. $A(f)$ is the aperture filter function and is defined by
\begin{equation}
A(f) = \mid \mathcal{F} (P(x,y)) \mid^2,
\label{eq:apfilter}
\end{equation}
where $p(r,\theta)$ is an arbitrary pupil function defined in cartesian coordinate system, $x$ and $y$. For a circular aperture, equal to one for $x^2+y^2 < D^2/4$ and zero elsewhere, $A(f) = (2\mathrm{J}_1(\pi D f)/(\pi D f))^2$ \citep{Tokovinin02}. $\phi$ is the frequency component of the refractive index power spectrum\comment{ (equation~\ref{eqn:scint_spectrum})}, for example, for Kolmogorov turbulence $\phi = f^{-11/3}$.

If the aperture cut-off frequency, $f_c \propto 1/D$ is small (i.e. telescopes with $D$ larger than a few tens of centimetres, $D\gg r_F$), such that $\pi \lambda z f_c^2 \ll 1$, we can invoke the small angle approximation and $S(z,f)$ can be written as $\left(\pi\lambda z f^2\right)^2/\lambda^2$. Introducing the dimensionless frequency $q=Df$, equation~\ref{eqn:scint_sigma_A} can be expressed by \citep{Kornilov12b},
\begin{equation}
\sigma_{\mathrm{se}}^2 = 38.44D^{-7/3}  \int C_n^2(z) z^{2} \mathrm{d}z \int \phi(q) q^3 (\mathrm{J}_1(\pi q))^2 \mathrm{d}q.
\label{eqn:Q_se_A}
\end{equation}

For exposure times longer than the speckle crossing time, $t_c = D/V_{\bot}$, where $V_{\bot}$ is the projected windspeed perpendicular to the optical axis, high frequency intensity variations are averged out, (i.e. the scintillation temporal power spectrum is filtered so that only low frequency components, $f<\pi D/V_{\bot}$ affect the scintillation power). The scintillation weighting function with an exposure time, $t$, larger than the crossing time of the speckles is then multiplied by the corresponding wind shear filter function $A_s = D/\pi t V_\bot q$ \citep{Kornilov12b}, and, 
\begin{equation}
\sigma_{\mathrm{le}}^2 = 12.24D^{-4/3}  \int \frac{C_n^2(z) z^{2}}{V_\bot(z)} \mathrm{d}z \int \phi(q) q^2 (\mathrm{J}_1(\pi q))^2 \mathrm{d}q.
\label{eqn:Q_le_A}
\end{equation}
The integrals over $q$ in the previous two equations are found by numerical integration to converge to 0.45 and 0.87 respectively. The scintillation index for short exposures becomes,
\begin{equation}
	\sigma_{I,\mathrm{se}}^{2} = 17.34D^{-7/3}\left(\cos\gamma\right)^{-3}\int_0^{\infty} h^{2}C_{n}^{2}\left(h\right)\mathrm{d}h,
\label{eq:scint_se_A}
\end{equation}
and for long exposures,
\begin{equation}
	\sigma_{I,\mathrm{le}}^{2} = 10.66D^{-4/3}t^{-1}\left(\cos\gamma\right)^{\alpha}\int_0^{\infty} \frac{ h^{2}C_{n}^{2}\left(h\right)}{V_{\bot}(h)} \mathrm{d}h,
	\label{eq:scint_le_A}
\end{equation}
where $h$ is the altitude of the turbulent layer, with $h=z\cos{(\gamma)}$, $\gamma$ being the zenith angle of the observation, $V_{\bot}(h)$ is the wind velocity profile and $\alpha$ is the exponent of the airmass. Note, that the value of the airmass exponent, $\alpha$, will depend on the wind direction and vary between $\left(\cos\gamma\right)^{-3}$ for the case when the wind is transverse to the azimuthal angle of the star, up to $\left(\cos\gamma\right)^{-4}$ in the case of a longitudinal wind direction. This difference comes from geometry, in the case where the wind direction is parallel to the azimuthal angle of the star the projected pupil onto a horizontal layer is stretched by a factor of $1/\cos\gamma$.

For small apertures, where the aperture size is smaller than the spatial scale of the intensity fluctuations ($D<r_F$), there is not enough spatial averaging to remove the dependance on the wavelength (the small angle approximation on the Fresnel filter function is no longer valid, $\sin^2\left(\pi\lambda z f^2\right)/\lambda^2 \ne (\pi\lambda z f^2)^2 /\lambda^2 $) and the scintillation index can be approximated by \citep{Dravins1998},
\begin{equation}
	\sigma_{I}^{2} = 19.2\lambda^{-7/6}\left(\cos\gamma\right)^{-11/6}\int_0^{\infty} h^{5/6}C_{n}^{2}\left(h\right)\mathrm{d}h.
	\label{eq:scint_smallD_A}
\end{equation}

If we replace the Kolmogorov power spectrum ($\phi = f^{-11/3}$) in equation~\ref{eqn:scint_weight_A} with a Von Karmen spectrum, modified to include a defined outer scale, $L_0$, of the form,
\begin{equation}
\phi(f) = \left(f^2 + L_0(h)^{-2}\right)^{-11/6}
\end{equation}
\comment{
\begin{equation}
\Phi_{\phi}(f) = 9.7 \times 10^{-3} k^{2}  \int_0^\infty C_n^{2}(h) \left(f^2 + L_0(h)^{-2}\right)^{-11/6}\mathrm{d}h,
\end{equation}
or in terms of $q$ in equation~\ref{eqn:Q_le_A},
\begin{equation}
\Phi_{\phi}(q) = 9.7 \times 10^{-3} k^{2}  \int_0^\infty C_n^{2}(h) \left(q^2 + (D/L_0(h)^{2})\right)^{-11/6}\mathrm{d}h,
\end{equation}
}
The integral in equations~\ref{eqn:Q_se_A} and \ref{eqn:Q_le_A} now have to be evaluated for the correct outer scale and become more complicated if the outer scale varies with altitude.

The scintillation index is now be expressed as,
\begin{multline}
\sigma_{\mathrm{le}}^2 = 12.24D^{-4/3} t^{-1}  \left(\cos\gamma\right)^{\alpha} \int \frac{C_n^2(h) h^{2}}{V_\bot(h)}  \\
\int \left(q^2 + (D/L_0(h)^{2})\right)^{-11/6} q^2 (\mathrm{J}_1(\pi q))^2 \mathrm{d}q \mathrm{d}h,
\label{eq:scint_gen_A}
\end{multline}
and can be solved numerically for the particular telescope and atmospheric parameter set in question.

With a central obscuration the aperture filter function can be expressed as \citep{Young1969},
\begin{equation}
A(f) = \frac{4}{\pi}\left[ \left(\frac{\mathrm{J}_1(\pi Df)}{\pi Df} \right) - \epsilon^2\left(\frac{\mathrm{J}_1(\pi \epsilon Df)}{\pi \epsilon Df} \right) / (1-\epsilon^2)\right]^2,
\label{eq:secondary_filter_A}
\end{equation}
where $\epsilon$ is the ratio of the diameter of the secondary obscuration to the diameter of the primary mirror. 

Equation~\ref{eq:secondary_filter_A} can be implemented in to the generalised scintillation index equation \comment{(equation~\ref{eqn:scint_weight_A})}by replacing the $(\mathrm{J}_1(\pi q))^2$ in equation~\ref{eq:scint_gen_A} with,
\begin{equation}
(\mathrm{J}_1(\pi q) - \epsilon \mathrm{J}_1(\pi \epsilon q))^2 / (1-\epsilon^2)^2.
\end{equation}

The final, generalised for outer scale and central obscuration, long exposure scintillation index is then given by,
\begin{multline}
\sigma_{\mathrm{le}}^2 = 12.24D^{-4/3} t^{-1}  \left(\cos\gamma\right)^{\alpha} \int \frac{C_n^2(h) h^{2}}{V_\bot(h)}  \\
\int \left(q^2 + (D/L_0(h)^{2})\right)^{-11/6} q^2 \\
(\mathrm{J}_1(\pi q) - \epsilon \mathrm{J}_1(\pi \epsilon q))^2 / (1-\epsilon^2)^2 \mathrm{d}q \mathrm{d}h,
\label{eq:scint_gen_long_A}
\end{multline}
and the generalised short exposure scintillation equation is,
\begin{multline}
\sigma_{\mathrm{se}}^2 = 38.44D^{-7/3}  \left(\cos\gamma\right)^{-3} \int C_n^2(h) h^{2}  \\
\int \left(q^2 + (D/L_0(h)^{2})\right)^{-11/6} q^3 \\
(\mathrm{J}_1(\pi q) - \epsilon \mathrm{J}_1(\pi \epsilon q))^2 / (1-\epsilon^2)^2 \mathrm{d}q \mathrm{d}h.
\label{eq:scint_gen_short_A}
\end{multline}

\end{document}